%% file: main.tex
\documentclass{INTERSPEECH2023}

\interspeechcameraready 
\usepackage{ieeetrantools}
\usepackage{booktabs}
\usepackage{multirow}
\usepackage{amssymb,amsmath} 
\usepackage{cite} 

\title{Adapting Multi-Lingual ASR Models for Handling Multiple Talkers}
\name{\begin{tabular}{c}
Chenda Li$^{1,2,\dag}$, Yao Qian$^2$, Zhuo Chen$^2$, Naoyuki Kanda$^2$,  \\
Dongmei Wang$^2$,
Takuya Yoshioka$^2$, Yanmin Qian$^1$, Michael Zeng$^2$
\end{tabular}  \thanks{$^{\dag}$The first author conducted this work during internship at Microsoft.}}
\address{
  $^1$Department of Computer Science and Engineering, Shanghai Jiao Tong University, Shanghai, China \\
  $^2$Microsoft, Redmond, WA, USA
  }
\email{$^1$\{lichenda1996, yanminqian\}@sjtu.edu.cn, \\
$^2$\{yaoqian, zhuc, nakanda, dongmei.wang, tayoshio, nzeng\}@microsoft.com}

\begin{document}
\bstctlcite{IEEEexample:BSTcontrol}

\maketitle
 
\begin{abstract}
State-of-the-art large-scale universal speech models (USMs) show a decent automatic speech recognition (ASR) performance across multiple domains and languages. However, it remains a challenge for these models to recognize overlapped speech, which is often seen in meeting conversations. We propose an approach to adapt USMs for multi-talker ASR. We first develop an enhanced version of serialized output training to jointly perform multi-talker ASR and utterance timestamp prediction. That is, we predict the ASR hypotheses for all speakers, count the speakers, and estimate the utterance timestamps at the same time. We further introduce a lightweight adapter module to maintain the multilingual property of the USMs even when we perform the adaptation with only a single language. Experimental results obtained using the AMI and AliMeeting corpora show that our proposed approach effectively transfers the USMs to a strong multilingual multi-talker ASR model with timestamp prediction capability. 
\end{abstract}
\input{tex/introduction.tex}

\input{tex/methods.tex}

\input{tex/exp.tex}

\section{Conclusion}

In this paper, we propose an effective approach to adapt USMs for multi-talker ASR. 
The enhanced version of serialized output training with timestamp prediction is first introduced for handling multiple-talkers with accurate speaker and time prediction. We also introduce the adapter module to transfer cross-lingual knowledge for multi-talker processing from a single language data. Our experiments show that the enhanced SOT successfully converts Whisper to multi-talker ASR
that can recognize overlapping speech while predicting the speaker and timestamps with high accuracy. 
By introducing the adapter module, our proposed model becomes able to handle multi-talker audio even when the language spoken in the audio is unseen in the adaptation data.

\bibliographystyle{IEEEtran}
\bibliography{mybib}

\end{document}

%% file: tex/introduction.tex
\section{Introduction}

Overlapping speech is omnipresent in natural speech communications~\cite{barkerFifthCHiMESpeech2018a,watanabeCHiME6ChallengeTackling2020a}. 
Previous studies showed that the duration of overlapped speech during meetings could range from 6\% to 15\%~\cite{cetinAnalysisOverlapsMeetings2006}.
Automatic conversation transcription systems must be able to recognize overlapping speech and distinguish individual speakers.
Additionally, it is desirable to estimate the timestamps of each utterance to enhance usability.

Single-talker automatic speech recognition (ASR) systems often struggle when they are presented with multi-talker overlapped speech \cite{qianSinglechannelMultitalkerSpeech2018,yoshiokaMeetingTranscriptionUsing2019}. 
Permutation invariant training (PIT) \cite{yuPermutationInvariantTraining2017,kolbaekMultitalkerSpeechSeparation2017} enables building ASR systems for multi-talker data \cite{yuRecognizingMultiTalkerSpeech2017,sekiPurelyEndtoEndSystem2018,changEndtoendMonauralMultispeaker2019,changMIMOSpeechEndtoEndMultiChannel2019,zhangImprovingEndtoEndSingleChannel2020}. 
Most of the PIT-based models have a fixed number of output branches, making them less useful for processing a long audio signal containing an unknown number of speakers. 
Target speech recognition~\cite{delcroixSingleChannelTarget2018,kandaAuxiliaryInterferenceSpeaker2019,huangAdaptingSelfsupervisedModels2022} is another approach, which only transcribes the speech of a specified talker.
Serialized output training (SOT) \cite{kandaSerializedOutputTraining2020a} was recently proposed to map the multi-talker audio including speech overlaps into a single output sequence representing the transcriptions of multiple speakers.
A special token indicating a speaker change is inserted in the output sequence to distinguish the transcriptions of individual speakers.
While the SOT-based ASR model can transcribe the overlapping speech of a variable number of speakers, the originally proposed scheme does not offer the ability to predict the utterance timestamps, limiting the usefulness of the generated transcriptions. 
It is also worth noting that SOT has only been tested in a monolingual setting.

The timestamp prediction of ASR hypothesis was conventionally achieved by using the forced alignment algorithm based on the phoneme posterior probability \cite{mcauliffe2017montreal}, which was common for hybrid ASR systems. 
However, 
it cannot be directly applicable
for end-to-end (E2E) ASR models \cite{li2022recent}
because the ASR model is usually not designed to 
produce phoneme posterior probability.
It becomes further challenging when the audio contains speech overlaps,
for which even a conventional forced alignment tool is no longer applicable.
Only limited studies were conducted to estimate timestamps of the multi-talker ASR hypothesis,
where a small neural network was introduced to estimate word timestamps \cite{kanda2022transcribe} 
or special symbols indicating the start and end of each utterance were introduced \cite{sklyar22_interspeech}.


 In this work, we aim to solve the 
multi-talker ASR, speaker counting, and timestamp prediction problem
with a single multilingual E2E model.
Thanks to the advanced deep learning architecture (e.g. \cite{vaswaniAttentionAllYou2017}) and the rapid growth of data and computing resources in recent years, large-scale pre-trained foundation models \cite{schneiderWav2vecUnsupervisedPretraining2019,baevskiWav2vecFrameworkSelfSupervised2020,hsuHuBERTSelfSupervisedSpeech2021,chenWavLMLargeScaleSelfSupervised2021,wangUniSpeechUnifiedSpeech2021,radfordRobustSpeechRecognition2022} have shown their strengths in many speech processing tasks \cite{yangSUPERBSpeechProcessing2021}, including speech separation \cite{huangInvestigatingSelfSupervisedLearning2022,songExploringWavLMSpeech2023}, speaker identification/diarization \cite{yangSUPERBSpeechProcessing2021,chenWavLMLargeScaleSelfSupervised2021,chenLargeScaleSelfSupervisedSpeech2022}, and speech recognition \cite{schneiderWav2vecUnsupervisedPretraining2019,baevskiWav2vecFrameworkSelfSupervised2020,hsuHuBERTSelfSupervisedSpeech2021,chenWavLMLargeScaleSelfSupervised2021,wangUniSpeechUnifiedSpeech2021,radfordRobustSpeechRecognition2022}.
This inspires us that, with proper transfer learning, a single large-scale multi-lingual pre-trained model could be extended to 
handle the multi-talker ASR, speaker counting, and timestamp prediction problems all at once.
With this perspective, 
we choose the recently proposed Whisper \cite{radfordRobustSpeechRecognition2022} as our foundation model, and fine-tune it by using an enhanced SOT framework that includes an utterance timestamp prediction task.
We also introduce
a lightweight adapter \cite{houlsbyParameterEfficientTransferLearning2019} to alleviate catastrophic forgetting of languages
that are not included in the fine-tuning data.
Based on the experiments using 
the AMI \cite{carlettaAMIMeetingCorpus2006} and AliMeeting \cite{yuM2MetIcassp20222022} corpora,
we demonstrate that the proposed framework provides a single multilingual model
that can recognize overlapping speech with timestamp prediction, while accurately counting the number of speakers.

%% file: tex/methods.tex
\begin{figure*}[t]

\centering
\includegraphics[width=0.82\linewidth]{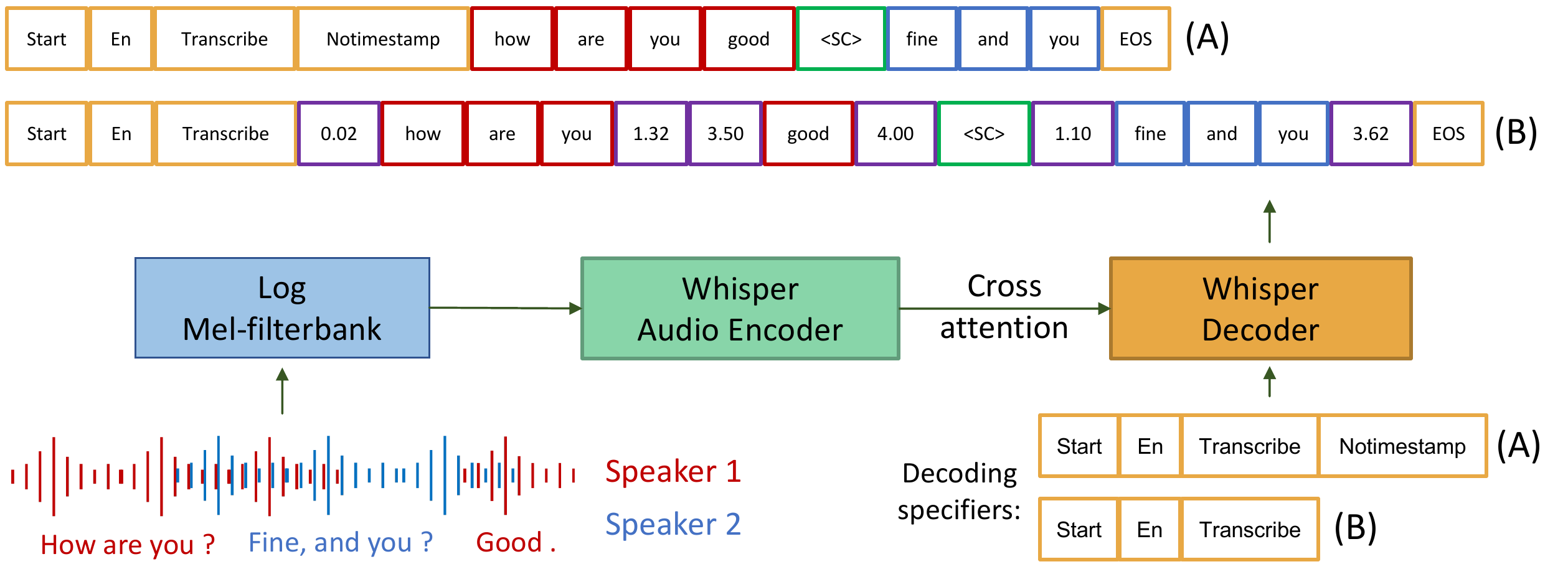}
\caption{An illustration of Whisper multi-talker transfer learning with SOT. (A): Training labels in SOT without timestamp; (B) Training labels in time-stamped SOT. }
\label{fig:fig_sot}

\end{figure*}


\section{Methods}

We leverage Whisper, one of the popular USMs, as a foundation model and adapt it to multi-talker speech recognition. The proposed enhanced SOT and adapter-based transfer learning methods are used to perform the adaptation. 

\subsection{Whisper as a foundation model}

Whisper is based on an attention-based encoder-decoder (AED) \cite{chorowskiAttentionBasedModelsSpeech2015,chanListenAttendSpell2016} architecture.
Both the audio encoder and text decoder comprise Transformer \cite{vaswaniAttentionAllYou2017} blocks.
The Whisper audio encoder encodes log mel-filterbank feature $\mathbf{X} \in \mathbb{R}^{f_{a}\times l_{a}}$ into deep hidden emebddings  $\mathbf{H} \in \mathbb{R}^{f_{h}\times l_{h}}$:
\begin{equation}
	\mathbf{H} = \text{Encoder}(\mathbf{X}),
\end{equation}
where $f_{a}, f_{h}$ and $l_{a}, l_{h}$ are the dimension and sequence length of input feature $X$ and hidden emebddings $H$, respectively.
The text decoder can be regarded as an audio-conditional autoregressive language model:
\begin{equation}
	\mathbf{o}_{n} =\text{Decoder}(\mathbf{t}, Y_{1:n-1} ,\mathbf{H}),
\end{equation}
where 
$\mathbf{o}_{n} \in \mathbb{R}^{|\mathcal{V}|}$ is an output distribution for estimating the $n$-th token $y_{n}$ from token vocabulary $\mathcal{V}$,
$Y_{1:N} = [y_1, \cdots , y_{N} | y_n \in \mathcal{V}]$ represents the estimated token sequence with length of $N$,
and $\mathbf{t} = \{t_{1}, \cdots, t_{I}| t_{i} \in \mathcal{V}\}$ is a sequence of special tokens\footnote{For example, 
$\mathbf{t}$ contains $[\langle \text{start} \rangle, \langle \text{English} \rangle, \langle \text{transcribe} \rangle, \langle \text{notimestamp} \rangle]$ for an English ASR task. Detailed descriptions can be found in  \cite[Sec. 2.3]{radfordRobustSpeechRecognition2022}.} that specifies a decoding task. 

Whisper was trained with multi-task learning on multilingual speech data. As a result, it showed remarkable multilingual ASR capabilities and thus can be seen as a USM. However, its ASR accuracy severely deteriorates when we apply it to a multi-talker speech recognition task. This could be because the multi-talker task was not explicitly included in the multi-task training.

\subsection{SOT with time-stamp prediction}
To address the multi-talker ASR challenge, we consider adapting a USM to build a single AED model that jointly performs multi-talker ASR, speaker counting, and timestamp prediction. To this end, we propose an enhanced version of SOT to incorporate the timestamp prediction task during the adaptation. 

SOT was proposed to train an AED model to perform multi-talker ASR \cite{kandaSerializedOutputTraining2020a}.
Although other multi-talker ASR models usually have multiple output branches \cite{yuPermutationInvariantTraining2017,yuRecognizingMultiTalkerSpeech2017,zhangImprovingEndtoEndSingleChannel2020}, SOT assumes an AED model with one output branch. Therefore, it can be used to convert a pre-trained USM into a multi-talker model without changing the model structure.
With SOT, the utterances of all speakers are concatenated to form a single token sequence by inserting a special token $\langle \text{sc} \rangle$ denoting a speaker change.
Fig.\ref{fig:fig_sot}.A illustrates how SOT organizes its training labels. 
The reference token sequence can be given as $R =[ r_{1}^{1}, \cdots, r^{1}_{N^1}, \langle \text{sc} \rangle, r_{1}^{2}, \cdots, r^{2}_{N^2},  \langle \text{eos} \rangle]$, where $r_{i}^{j}$ is the $i$-th token of the $j$-th speaker and $\langle \text{eos} \rangle$ denotes the end of the output sequence that is only used when the utterances of all speakers are transcribed.
We use the speaker-wise first-in-first-out (FIFO) style \cite{kandaSerializedOutputTraining2020a} to form the training labels.
That is, the reference transcriptions of individual speakers are sorted by the start times of their first utterances. 
As a consequence, 
the model is expected to transcribe all the utterances of the first speaker, then output a $\langle \text{sc} \rangle$ token, and then continue transcribing the remaining speakers one by one. 

We propose to incorporate the timestamp prediction task into SOT by leveraging 
the timestamp tokens of Whisper \cite{radfordRobustSpeechRecognition2022}.
Whisper includes a set of additional tokens in the token vocabulary $\mathcal{V}$ to 
represent discretized timestamps at a $20$-ms resolution.
These timestamp tokens are predicted by the decoder at the same time as the tokens representing the transcriptions.
We incorporate the timestamp tokens into SOT
as illustrated in Fig. \ref{fig:fig_sot}.B. 
During training, the timestamp tokens are inserted in the training labels at the beginning and the ending of each speaker's homogeneous segment.
The speaker homogeneous segment consists of consecutive tokens spoken by the same speaker where the silence periods between the neighboring tokens do not exceed $2$ seconds. 
Note that, unlike the case with Whisper, the timestamps do not increase monotonically within the token sequence and have overlapped regions among different speakers. 

\begin{table*}[th]
      \caption{WER (\%) and LDER (\%) for AMI-SDM evaluation set. }
      \vspace{-3mm}
  \label{tab:ami_sdm}
  \centering
\setlength{\tabcolsep}{1.1mm}{
\footnotesize
\begin{tabular}{@{}c|cccc|ccccc|ccccc}
\toprule
\multirow{2}{*}{\begin{tabular}[c]{@{}c@{}}Exp. \\ ID\end{tabular}} & \multirow{2}{*}{\begin{tabular}[c]{@{}c@{}}Pre-training \\ model\end{tabular}} & \multirow{2}{*}{\begin{tabular}[c]{@{}c@{}}Pre-training \\ data\end{tabular}} &\multirow{2}{*}{\begin{tabular}[c]{@{}c@{}}Fine-tuning \\ data\end{tabular}} & \multirow{2}{*}{\begin{tabular}[c]{@{}c@{}}Timestamp\\ prediction\end{tabular}} &  \multicolumn{5}{c|}{WER (w.r.t. \# of talkers) (\%)}  & \multicolumn{5}{c}{LDER (w.r.t. \# of talkers) (\%)}\\ 
& & & && avg. & 1 & 2 & 3 & 4 & avg. & 1 & 2 & 3 & 4 \\
\midrule
 - &   \begin{tabular}[c]{@{}c@{}}Multi-talker \\ AED \cite{kandaLargeScalePreTrainingEndtoEnd2021b} \end{tabular} & \begin{tabular}[c]{@{}c@{}} 900k hrs multi-talker\\ simulation \end{tabular}     &  AMI-SDM           &    -    &  21.2  & 14.7 & 19.6 & 25.7 & 35.5 &  \multicolumn{5}{c}{-}\\ 
\midrule
1       & \multirow{3}{*}{\begin{tabular}[c]{@{}c@{}}Whisper \\ small\end{tabular}}     &    \multirow{3}{*}{\begin{tabular}[c]{@{}c@{}}680k hrs\\Internet\end{tabular}}   &  -  &    -    &  47.3 & 25.4 & 46.3 & 68.0 & 82.5 & \multicolumn{5}{c}{-}  \\ 
2       &   &   & AMI-SDM     &    w/o     & 32.5 & 14.9 & 27.0 & 45.7 &  68.9 & \multicolumn{5}{c}{-} \\ 
3       &   &   & AMI-SDM     &    w       &  28.5  & 14.7 & 25.9 & 41.2  & 54.7 & 10.4 & 0.9 &6.8 & 21.7 & 35.3   \\ \midrule

4        & \multirow{3}{*}{\begin{tabular}[c]{@{}c@{}}Whisper \\ medium\end{tabular}}     &    \multirow{3}{*}{\begin{tabular}[c]{@{}c@{}}680k hrs\\Internet\end{tabular}}   & -           &    -       & 44.1 & 21.1 &  43.4 & 65.4  & 78.6 & \multicolumn{5}{c}{-}  \\ 
5       &   &     & AMI-SDM     &    w/o       & 26.8 & 12.3 & 21.6 & 36.8 & 62.2 & \multicolumn{5}{c}{-} \\ 
6       &    &  & AMI-SDM     &            w   & 23.6 & 12.8 & 21.8 & 32.5 & 45.9 & 8.2 & 0.9 & 5.9 & 16.5 & 27.6  \\ \midrule
7       & \multirow{3}{*}{\begin{tabular}[c]{@{}c@{}}Whisper \\ large\end{tabular}}     &    \multirow{3}{*}{\begin{tabular}[c]{@{}c@{}}680k hrs\\Internet\end{tabular}}   & -           &    -         & 44.2 & 22.5 & 44.6  &  65.0 & 78.5 & \multicolumn{5}{c}{-} \\ 
8       &   &   & AMI-SDM     &    w/o      & 28.4 & 13.0 & 23.0 & 38.7 & 66.0 & \multicolumn{5}{c}{-} \\ 
9       &   &  & AMI-SDM     &            w & 21.4  & 12.0 &  20.0 & 29.3 & 40.6  & 6.2 & 0.8 & 4.5 & 11.5 & 22.3 \\ 

\bottomrule
\end{tabular}
}
      \vspace{-3mm}
\end{table*}

\subsection{Adapter for cross-lingual transfer learning}

Whisper was pre-trained on $680$K hours of multilingual audio, of which a $117$K hours subset covered $96$  non-English languages \cite{radfordRobustSpeechRecognition2022}. 
It is desirable if a model adapted with multi-talker English data can also recognize the multi-talker speech of other languages that are not seen during the adaptation. 
In our preliminary experiments, when we fine-tuned the entire Whisper model with the enhanced SOT method using English data, 
the obtained model performed poorly for other languages. 

To improve multilingual generalization,
we employ a parameter-efficient transfer learning method proposed in  \cite{houlsbyParameterEfficientTransferLearning2019}.
Specifically, we insert bottleneck adapter modules 
with a limited number of parameters into the Transformer basic blocks of both the encoder and decoder.
We optimize only the newly introduced parameters and freeze most of the original parameters with the aim that the newly added parameters could learn to handle multiple talkers in a less language-dependent fashion.
The detailed structure of the adapter module will be described in Sec.\ref{sec:train_cfg}.
 

%% file: tex/exp.tex
\section{Experiments}
\subsection{Dataset}

\subsubsection{English Data}
We use the AMI meeting corpus \cite{carlettaAMIMeetingCorpus2006}, which contains approximately 100 hours of English meeting recordings, for both SOT-based fine-tuning and testing.
We utilize the audio signals from the first channel of the microphone array, a.k.a. the single distant microphone (SDM) audio. 
We segment each recording at silence positions or non-overlapping utterance boundaries into shorter \textit{utterance groups} for both training and evaluation. 
Further details of \textit{utterance groups} and the dataset specification can be found in \cite{kandaLargeScalePreTrainingEndtoEnd2021b}.
After the segmentation, there is a total of 66 hours of SDM training data. 

We also utilize 120-hour or 360-hour meeting-style English data simulated using LibriSpeech \cite{panayotovLibrispeechASRCorpus2015} in some experiments. 
Our simulation setup follows the one used in \cite{liDualPathModelingLong2021,liDualPathModelingMemory2022}, except that we use \textit{utterance groups} of $30$ seconds or shorter and that there are at most $4$ speakers in each \textit{utterance group}.
To promote the learning of multi-talker recognition, 
these simulated data have more utterance overlaps than the AMI-SDM segments. The overlap ratio of each \textit{utterance group} is between $60\%$ and $80\%$.

\subsubsection{Mandarin Chinese Data}
We use AliMeeting \cite{yuM2MetIcassp20222022} to test our fine-tuned multi-talker models in different languages with a zero-shot approach.
The AliMeeting corpus contains approximately 108 hours of real meeting recordings in Mandarin Chinese.
We use the audio signals from the third channel of the far-field microphone array.
The meeting-long recordings are segmented into \textit{utterance groups} in the same manner as with AMI-SDM.
The \textit{utterance groups} that are longer than $30$ seconds are not used. 
As a result, there are $31.3$, $1.5$, and $3.9$ hours of data for training, validation, and testing, respectively.
The training set is only used in Exp. 10 of Table \ref{tab:alimeeting} to provide a reference point for the cross-lingual task. 

\subsection{Evaluation}

At inference time, the multi-talker fine-tuned models generate utterances for one or multiple speakers. 
For the English test set, we calculate word error rate (WER) by following the procedure described in \cite{kandaLargeScalePreTrainingEndtoEnd2021b}.
Namely, for each utterance group, 
the best speaker alignment between the hypotheses and the references 
is selected among all possible permutations.
The WER is calculated by dividing the total error count from all utterance groups by the total number of reference words.
 For the Mandarin test set, we compute the character error rate (CER) by using the same procedure
 except that we use the characters as the recognition unit.
The text normalization scheme of  Kaldi \cite{poveyKaldiSpeechRecognition2011} AMI recipe is applied in the AMI-SDM evaluation.
For AliMeeting, we use Whisper's default text normalizer and then convert all the traditional Chinese characters into simplified Chinese\footnote{A  traditional Chinese character and its simplified version differ in writing but have the same pronunciation and meaning.}.

In addition, to assess the quality of speaker and timestamp prediction, we compute the diarization error rate (DER) within each \textit{utterance group}, which we call local DER (LDER). 
We also report the speaker counting accuracy for each \textit{utterance group}
by following \cite{kandaLargeScalePreTrainingEndtoEnd2021b}.

\begin{table}[th]
  \caption{AMI-SDM speaker counting accuracy comparison between SOT baseline (Exp. 5) and joint timestamp prediction (Exp. 6).}
      \vspace{-3mm}
  \label{tab:ami_sdm_acc}
  \centering
\setlength{\tabcolsep}{1.3mm}{
\footnotesize
\begin{tabular}{@{}c|c|rrrrr}
\toprule
\multirow{2}{*}{Exp. ID} & \multirow{2}{*}{\begin{tabular}[c]{@{}l@{}}Actual \# \\ of talkers\end{tabular}} & \multicolumn{5}{c}{Estimated \# of talkers (\%)} \\
                         &     & 1     & 2    & 3    & 4    & $\geq$5    \\ \midrule
\multirow{4}{*}{5}        & 1  &   \textbf{98.6}    &   1.4  &   0.1   &  0.0    &  0.0   \\
                         & 2   &    16.9   &   \textbf{76.2}  &   6.8  &   0.2   & 0.0   \\
                         & 3   &   4.0    &  40.9    &  \textbf{46.6}    &  7.9    & 0.6   \\
                         & 4   &   1.5    &  17.7    &  42.9   & \textbf{28.6}   &  9.3  \\ \midrule
\multirow{4}{*}{6}        & 1  &   \textbf{97.7}    &  2.1    &  0.2   &  0.0  &  0.0  \\
                         & 2   &  12.6     &   \textbf{72.3}   &  14.0    &  1.0  & 0.0   \\
                         & 3   &   1.6    &  24.1    &  \textbf{56.0}    &  15.9   &   2.3 \\
                         & 4   &   0.0    &   8.2  &  39.2   &   \textbf{35.6}  &   16.9 \\  
                         \bottomrule
\end{tabular}
}
\vspace{-3mm}
\end{table}

\begin{table*}[th]
  \caption{Comparison of generalization ability for unseen language (Mandarin Chinese) in multi-talker processing.}
  \label{tab:alimeeting}
\vspace{-3mm}
  \centering
\setlength{\tabcolsep}{1.5mm}{
\footnotesize
\begin{tabular}{c|cc|cccccc|cccccc}
\toprule
\multirow{3}{*}{\begin{tabular}[c]{@{}c@{}}Exp.\\ Id\end{tabular}} & \multirow{3}{*}{\begin{tabular}[c]{@{}c@{}}Fine-tuning\\ data\end{tabular}} & \multirow{3}{*}{Adapter} & \multicolumn{6}{c|}{AMI-SDM  (English)} & \multicolumn{6}{c}{AliMeeting (Mandarin)} \\ \cline{4-15} 
 &   &  & \multicolumn{1}{c|}{\multirow{2}{*}{WER}} & \multicolumn{1}{c|}{\multirow{2}{*}{LDER}} & \multicolumn{4}{c|}{SC Acc in \# talkers} & \multicolumn{1}{c|}{\multirow{2}{*}{CER}} & \multicolumn{1}{c|}{\multirow{2}{*}{LDER}} & \multicolumn{4}{c}{SC Acc in \# talkers} \\
 &   &  & \multicolumn{1}{c|}{} & \multicolumn{1}{c|}{} & 1 & 2 & 3 & 4 & \multicolumn{1}{c|}{} & \multicolumn{1}{c|}{} & 1 & 2 & 3 & 4 \\ \midrule
4  & - & - & \multicolumn{1}{c|}{44.0} & \multicolumn{1}{c|}{-} & 100.0 & 0.0 & 0.0 & 0.0 & \multicolumn{1}{c|}{59.1} & \multicolumn{1}{c|}{-} & 100.0 & 0.0 & 0.0 & 0.0 \\
\midrule
10 & AliMeeting & w/o & \multicolumn{1}{c|}{39.4} & \multicolumn{1}{c|}{22.7} & 99.3 & 12.3 & 1.5& 0.5 & \multicolumn{1}{c|}{28.3} & \multicolumn{1}{c|}{12.8} &95.3 & 80.2 & 52.6 & 28.8 \\ 
\midrule
6 & AMI-SDM  & w/o & \multicolumn{1}{c|}{23.6} & \multicolumn{1}{c|}{8.2} & 97.7 & 72.3 & 56.0 & 35.6 & \multicolumn{1}{c|}{50.7} & \multicolumn{1}{c|}{35.2} & 91.8 &  65.4 & 26.9& 11.1  \\
11 & AMI-SDM  & w & \multicolumn{1}{c|}{25.0} & \multicolumn{1}{c|}{9.3} & 97.6 & 73.9 & 46.5 & 29.3 & \multicolumn{1}{c|}{43.7} & \multicolumn{1}{c|}{26.9} & 95.7  &  60.4 & 21.0 & 5.2 \\
12 & \quad + LibriSpeech 120h  & w & \multicolumn{1}{c|}{24.7} & \multicolumn{1}{c|}{9.0} & 97.4 & 74.4 & 49.3 & 28.7 & \multicolumn{1}{c|}{40.1} & \multicolumn{1}{c|}{22.1} & 89.1  &  72.3 & 32.9 & 20.0 \\
13 & \quad + LibriSpeech 360h  & w & \multicolumn{1}{c|}{24.3} & \multicolumn{1}{c|}{8.7} & 98.4 & 73.9 & 47.3 &  27.1 & \multicolumn{1}{c|}{36.9} & \multicolumn{1}{c|}{18.0} & 91.9  & 73.2 & 28.7 & 16.3  \\

\bottomrule
\end{tabular}
}
\vspace{-3mm}
\end{table*}

\subsection{Training configuration}
\label{sec:train_cfg}

We fine-tune Whisper models based on the AMI-SDM training data with the enhanced SOT-style reference transcriptions.
Three Whisper models, including small, medium, and large (v1) \cite{radfordRobustSpeechRecognition2022}, are examined.
We conduct fine-tuning experiments without and with the adapter module.

For the experiments without the adapter module,
we first extend the token embedding set of the decoder by adding a randomly-initialized embedding representing
$\langle \text{sc} \rangle$ token.
All the Whisper model parameters are then updated with
the AdamW optimizer.
We use a linear decay learning rate (LR) scheduler with the initial LR of $1e-6$ (Exp. \{2,3,5,6,8,9,10\}).
We fine-tune the small and medium models for $2$ epochs with a batch size of $2$ while a batch size of $1$ is used for the large model.

For the experiments using the adapter module, we freeze most of the original Whisper model parameters, 
and only update the parameters in the layer normalization modules in the Transformer blocks, the token embedding for $\langle \text{sc} \rangle$, and the parameters constituting the adapter module.
The initial LR of 1e-5 is used.
The adapter module is a two-layer bottleneck feed-forward network with a ReLU activation function. The input and output dimensions are the same as 
the Whisper model's width
and the bottleneck dimension is $256$.
We add two adapter modules to each Transformer block contained in the Whisper encoder and decoder: one after the self-attention layer and one after the feed-forward layer.
There is a residual connection \cite{heDeepResidualLearning2016} within each adapter module, and near-identity initialization \cite{houlsbyParameterEfficientTransferLearning2019} is applied to the adapter modules at the start of fine-tuning.

\subsection{Results on AMI-SDM}

We first evaluate models fine-tuned without adapter modules by using the AMI-SDM evaluation set. 
Table \ref{tab:ami_sdm} shows the WER comparison of the original Whisper models (Exp. \{1,4,7\}), the models fine-tuned with the original SOT without timestamp prediction (Exp. \{2,5,8\}) and the models fine-tuned with the enhanced SOT (Exp. \{3,6,9\}).
Large WER margins between the original models and the SOT fine-tuned models are consistently observed for all three model sizes.  This shows the SOT fine-tuning makes the Whisper models capable of handling the overlapped multi-talker speech.
By comparing the experiments with/without timestamp prediction, we find that the models fine-tuned with timestamp prediction greatly reduce the WERs for the utterance groups including 3- and 4-speakers.
We further calculate the speaker counting accuracy and report it as a confusion matrix in Table \ref{tab:ami_sdm_acc}.
From the result, we can observe that the addition of the timestamp prediction task helps improve the speaker counting accuracy for the utterance groups with 3 and 4 speakers, which aligns with the significant WER improvements.
Our hypothesis for these improvements is that the timestamp tokens can serve as an explicit indicator for the SOT decoder to identify the location of the overlapping region.
As a result, the learning of the cross-attention in the decoder may have been facilitated.
Further investigation is desirable to better understand this phenomenon, and we will leave it to future work.

We also include the state-of-the-art (SOTA) system on the same testing setup (i.e. utterance-group-level evaluation that does not allow the use of oracle utterance boundaries) from previous work \cite{kandaLargeScalePreTrainingEndtoEnd2021b} for better comparison.
Our best system (Exp. 9) achieves very close overall WER to the SOTA results but there is still a gap for the $3$- and $4$-speakers breakdown WER.
The reason might be that large-scale simulated multi-talker speech is used for SOT pre-training in \cite{kandaLargeScalePreTrainingEndtoEnd2021b}, and their pre-trained model already has some good knowledge for multi-talker processing.

The LDERs of the enhanced SOT-based systems are also listed in Table \ref{tab:ami_sdm}. 
We can see a clear trend that the LDER becomes better as the model size increases, especially for the utterance groups with many speakers.
These results show that the obtained model can perform ASR and timestamp prediction for multiple talkers while counting them at the same time.

\subsection{Results on Zero-shot Multilingual Transfer}
\label{sec:alimeeting}

Table \ref{tab:alimeeting} reports the WERs and LDERs of different systems on the AliMeeting testset to see if the models adapted by the English multi-talker data are applicable to the Mandarin testset. 
We first evaluate the results on the original Whisper medium model (Exp. 4) and the AliMeeting multi-talker fine-tuned model (Exp. 10). These results can be regarded as the lower- and upper-bound for other evaluations in the table.

The model without the adapter module (Exp. 6) achieves the best performance on the AMI-SDM dataset, but it underperforms all the other fine-tuned models on the AliMeeting dataset.
It shows that naively fine-tuning the model with the English-only multi-talker data impairs the multi-lingual capability which the original model is equipped with. 

On the other hand, we observe significant improvements in both WER and LDER for the AliMeeting testset
when we introduce the adapter module, while mostly preserving the same WER and LDER for the AMI-SDM testset (Exp. 6 vs Exp. 11).
This result shows that the adapter module can help the model acquire the multi-talker recognition capability in a less language-dependent way so that the obtained model can handle the multi-talker data in multiple languages. 

We further fine-tune the model with additional simulated LibriSpeech meeting-style data with high overlap ratios. 
Comparing Exp. \{12,13\} with Exp. 11, there are large improvements in speaker counting accuracy for the 3- and 4-speaker utterance groups.
It shows that using more overlapped English training data facilitates the learning of multi-talker processing, and the learned knowledge from English can be well transferred into Chinese multi-talker meetings.